\documentclass[conference]{IEEEtran}
\IEEEoverridecommandlockouts
\usepackage{cite}
\usepackage{amsmath,amssymb,amsfonts}
\usepackage{algorithmic}
\usepackage{graphicx}
\usepackage{textcomp}
\usepackage{xcolor}

\usepackage[linesnumbered,ruled,vlined]{algorithm2e}

\SetCommentSty{mycommfont}
\SetKwInput{KwInput}{Input}                % Set the Input
\SetKwInput{KwOutput}{Output} 

\def\BibTeX{{\rm B\kern-.05em{\sc i\kern-.025em b}\kern-.08em
    T\kern-.1667em\lower.7ex\hbox{E}\kern-.125emX}}
\begin{document}

\title{Improving Performance in Combinatorial Optimization Problems with Inequality Constraints: An Evaluation of the Unbalanced Penalization Method on D-Wave Advantage\\
\thanks{This work has been submitted to the IEEE for possible publication. Copyright may be transferred without notice, after which this version may no longer be accessible.}
}

\author{\IEEEauthorblockN{J. A. Monta\~nez-Barrera\IEEEauthorrefmark{1}, Pim van den Heuvel\IEEEauthorrefmark{1}, Dennis Willsch\IEEEauthorrefmark{1}, Kristel Michielsen\IEEEauthorrefmark{1}\IEEEauthorrefmark{2}\IEEEauthorrefmark{3}}
\IEEEauthorblockA{\textit{\IEEEauthorrefmark{1}Jülich Supercomputing Centre} \\
\textit{Institute for Advanced Simulation, Forschungszentrum Jülich}\\
52425 Jülich, Germany}
\IEEEauthorblockA{\textit{\IEEEauthorrefmark{2}AIDAS}\\
52425 Jülich, Germany}
\IEEEauthorblockA{\textit{\IEEEauthorrefmark{3}RWTH Aachen University} \\
52056 Aachen, Germany\\
j.montanez-barrera@fz-juelich.de}
}

\maketitle

\begin{abstract}
Combinatorial optimization problems are one of the target applications of current quantum technology, mainly because of their industrial relevance, the difficulty of solving large instances of them classically, and their equivalence to Ising Hamiltonians using the quadratic unconstrained binary optimization (QUBO) formulation. Many of these applications have inequality constraints, usually encoded as penalization terms in the QUBO formulation using additional variables known as slack variables. The slack variables have two disadvantages: (i) these variables extend the search space of optimal and suboptimal solutions, and (ii) the variables add extra qubits and connections to the quantum algorithm. Recently, a new method known as unbalanced penalization has been presented to avoid using slack variables. This method offers a trade-off between additional slack variables to ensure that the optimal solution is given by the ground state of the Ising Hamiltonian, and using an unbalanced heuristic function to penalize the region where the inequality constraint is violated with the only certainty that the optimal solution will be in the vicinity of the ground state. This work tests the unbalanced penalization method using real quantum hardware on D-Wave Advantage for the traveling salesman problem (TSP). The results show that the unbalanced penalization method outperforms the solutions found using slack variables and sets a new record for the largest TSP solved with quantum technology.
\end{abstract}


\section{Introduction}

Combinatorial optimization is an important area of research because it provides efficient algorithms and tools to solve a wide range of problems that arise in different fields, such as computer science \cite{Applegate2006, Tomita2007}, finance \cite{Cornuejols2018, Mugel2022}, logistics \cite{dantzig1959truck, Wang2010, Gao2020}, operations research \cite{OperationsResearch}, and biology \cite{klepeis2003astro}. Many real-world problems can be modeled as combinatorial optimization problems, where the goal is to find the best possible solution from a finite set of possible solutions. 

Over the past few years, there has been significant interest in exploring the potential of quantum computing to enhance the solving of combinatorial optimization problems. This is driven by several factors. Firstly, Ising Hamiltonians can be used to encode combinatorial optimization problems, with the ground state representing the optimal solution of the problem \cite{Lucas2014, Kochenberger2014}. Secondly, such problems are typically challenging to solve and have practical applications \cite{Ohzeki2020}. Thirdly, quantum algorithms designed to tackle these problems require minimal resources and can be tested on current state-of-the-art quantum hardware \cite{Harrigan2021, Niroula2022}. Finally, they can be used to benchmark the capabilities of new quantum algorithms and technologies \cite{Harrigan2021, GonzalezCalaza2021, Willsch2022 ,Lubinski2023}.

The standard method for encoding combinatorial optimization problems on quantum processing units (QPUs) is to transform them into their quadratic unconstrained binary optimization (QUBO) representation and obtain the Ising Hamiltonian after a change of variables. In the QUBO encoding, the constraints of combinatorial optimization problems are added to the cost function as penalization terms alongside the objective function. However, many of these problems rely on inequality constraints to define feasible solutions, and the traditional approach is to use slack variables. To address some of the limitations of slack variables, a new method called unbalanced penalization has been proposed \cite{Montanez-Barrera2022}. This method uses an unbalanced function to distinguish between feasible and infeasible regions of the inequality constraint by adding large penalization when the condition is not fulfilled. 

In this work, we investigate the effectiveness of the unbalanced penalization method for encoding inequality constraints in combinatorial optimization problems, and compare its performance with that of the slack variables approach. Specifically, we focus on the Dantzig-Fulkerson-Johnson (DFJ) \cite{dantzig1954solution} formulation of the Traveling Salesman Problem (TSP), which is a linear programming (LP) representation, and use random instances of the TSP with varying number of cities (ranging from 6 to 45). Then, we use a sub-tour elimination method that iteratively adds inequality constraints when they appear in the solution and evaluate its performance using quantum annealing and classical solvers. Our findings indicate that the unbalanced penalization method outperforms the slack variables approach, yielding better results on both quantum annealing and classical solvers.

The rest of the paper is organized as follows. Section \ref{Sec:Methods} provides a description of the implementation of the combinatorial optimization problems using the QUBO formulation, an overview of the slack variables and unbalanced penalization encodings, the Ising Hamiltonian encoding, and the formulation of the TSP for sub-tour elimination. Section \ref{Sec:Results} presents results of the unbalanced penalization approach and those of the slack variables approach using D-Wave Advantage, the D-Wave hybrid solver, and different classical solvers. Finally, Section \ref{Sec:Conclusions} provides some conclusions.

\section{Methods}\label{Sec:Methods}
\subsection{The QUBO formulation}\label{Sec:QUBO}
The QUBO formulation is a way to express a combinatorial optimization problem in the form of a quadratic objective function that depends on binary variables. In this formulation, the goal is to find the values of the binary variables that minimize the objective function subject to certain constraints. The set of combinatorial problems that can be represented by the QUBO formulation are characterized by functions of the form

\begin{equation}
f(\mathrm{x}) = \sum_{i=0}^{n-1} \sum_{j=0}^{n-1} q_{ij} x_{i} x_{j}, 
\end{equation}
where $n$ is the number of variables, $q_{ij} \in \mathbb{R}$ are coefficients associated to the specific problem, and $x_i \in \{0,1\}$ are the binary variables of the problem. Note that $x_{i} x_{i} \equiv x_{i}$ and $q_{ij} = q_{ji}$ in this formulation. 

The general form of a combinatorial optimization problem amenable to solution by QPUs is given by the cost function

\begin{equation}\label{QUBO_form}
f(\mathrm{x}) = 2\sum_{i=0}^{n-1} \sum_{j > i}^{n-1} q_{ij}x_{i}x_{j} + \sum_{i=0}^{n-1} q_{ii} x_i,
\end{equation}
and equality constraints given by

\begin{equation}
\sum_i c_i x_i = C, \ c_i \in \mathbb{Z},
\end{equation}
and inequality constraints given by

\begin{equation}\label{Eq:ineq}
 \sum_i w_i x_i \le W, \ w_i \in \mathbb{Z}.
\end{equation}
To transform problems with constraints into the QUBO formulation, the constraints are added as penalization terms. To this end, the equality constraints are included in the cost function using the penalization term

\begin{equation}\label{EQ_F}
\lambda_0 \left(\sum_i c_i x_i - C\right)^2,
\end{equation}
where $\lambda_0$ is a penalization coefficient that should be chosen appropriately to find enough solutions where the equality constraint is fulfilled (see Section \ref{Sec:Results} below) and $C$ is a constant value.
In the case of the inequality constraints we compare two methods, namely the unbalanced penalization presented in Sec. \ref{Sec:unbalanced} and the slack encoding presented in Sec. \ref{slack}. Any further equality or inequality constraints can be incorporated into the QUBO formulation by following the same principles outlined here and in the following subsections.

\subsection{Unbalanced penalization}\label{Sec:unbalanced}
The unbalanced penalization \cite{Montanez-Barrera2022} is a heuristic method for including inequality constraints as penalization terms in the QUBO formulation of combinatorial optimization problems. Starting from Eq. (\ref{Eq:ineq}), the method adds a penalization term $\xi(\mathrm{x})$ to the objective function given by

\begin{equation}
\xi(\mathrm{x}) = -\lambda_1 h(\mathrm{x}) + \lambda_2 h(\mathrm{x})^2
\label{eq:xi}
\end{equation}
where $h(\mathrm{x}) = W - \sum_i w_i x_i$ and $\lambda_{1,2}$ are penalization coefficients that should be chosen to guarantee that the constraint is fulfilled. Note that $h(\mathrm{x}) \ge 0$ due to the constraint in Eq. (\ref{Eq:ineq}). Similarly, a formulation can be devised for the scenario where $h(\mathrm{x}) \le 0$. The term $\xi(\mathrm{x})$ is unbalanced, meaning that it imposes a larger penalization for negative values of $h(\mathrm{x})$ (i.e., when the constraint is not satisfied) than for positive values. The QUBO formulation using the unbalanced penalization approach is given by 

 \begin{eqnarray}\label{QUBO_unbalanced}
 \min_x \left(2\sum_{i=0, j>i}^{n-1} q_{ij}x_{i}x_{j} + \sum_{i=0}^{n-1} q_{ii} x_i 
 + \lambda_0 \left(\sum_{i=1}^n c_i x_i - C\right)^2\nonumber\right.\\
\left.- \lambda_1\left(W - \sum_i w_i x_i\right) +  \lambda_2 \left(W - \sum_i w_i x_i\right)^2 \right).
 \end{eqnarray}

\begin{figure}[!htbp]
\includegraphics[width=8.5cm]{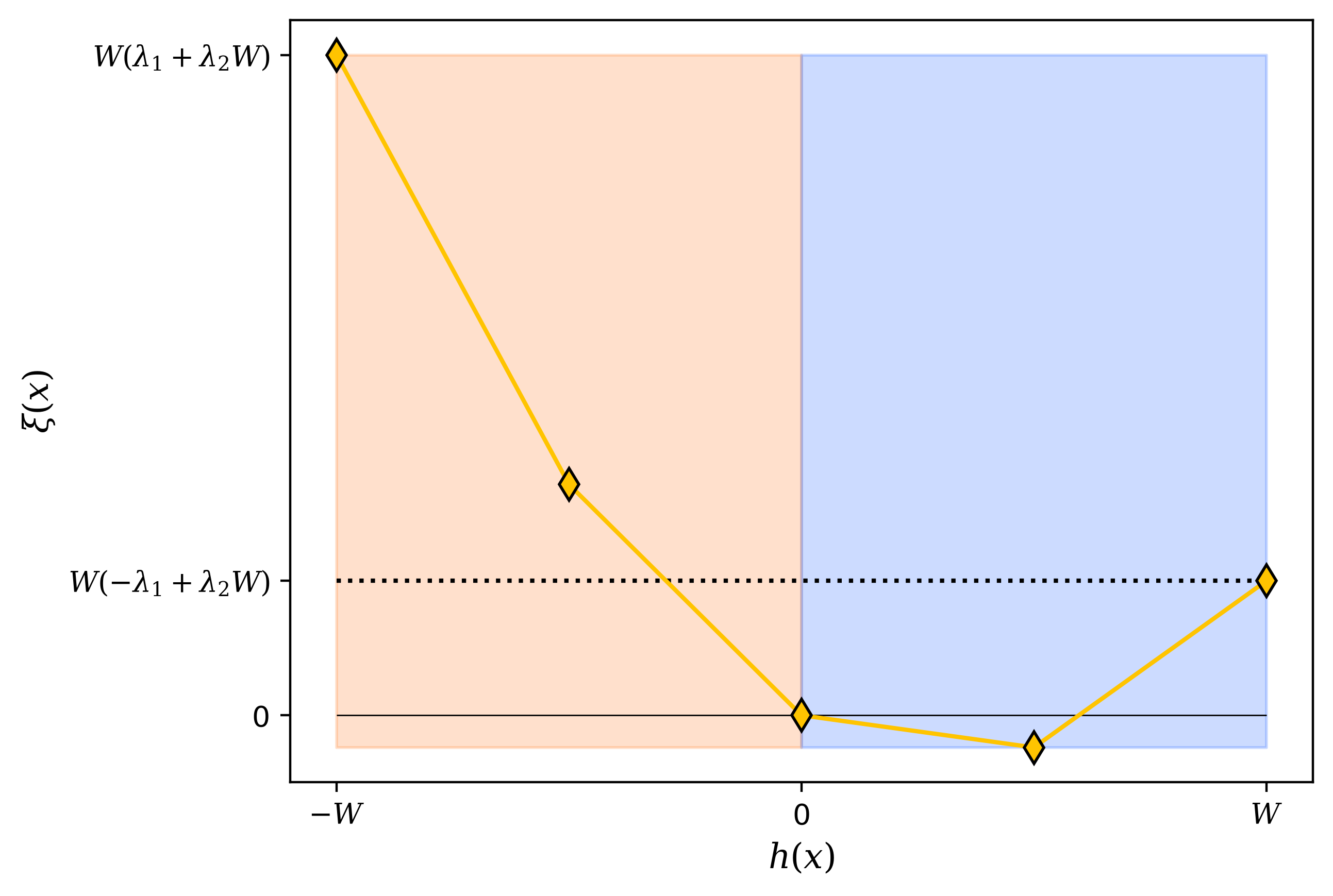}
\caption{Unbalanced penalization function $\xi(\mathrm{x})$ in Eq.~(\ref{eq:xi}) as a function of $h(x) = W - \sum_i w_i x_i$. The region $h(x)\ge0$ where the inequality constraint Eq.~(\ref{Eq:ineq}) is satisfied (violated) is shown in blue (orange). Markers represent special points at $h(x)\in\{-W,-W/2,0,W/2,W\}$.}
\label{unbalanced}
\end{figure}

Figure \ref{unbalanced} illustrates the unbalanced penalization function $\xi(\mathrm{x})$, where the blue region represents the region where the inequality constraint is satisfied, and the orange region represents the region where the inequality constraint is not satisfied, and therefore, adds a large penalization to the objective function.

\subsection{Slack variables}\label{slack}

The most common approach for including an inequality constraint in the QUBO formulation is to use a slack variable $S$ \cite{Glover2019}. The slack variable is introduced as an auxiliary variable that makes a penalization term vanish when the inequality constraint is satisfied. Specifically, if the inequality constraint of Eq. (\ref{Eq:ineq}) is given then the slack variable $S$ can be added as

\begin{equation}\label{ineq2}
W - \sum_i w_i x_i - S = 0.
\end{equation}
Note that when the inequality constraint in Eq. (\ref{Eq:ineq}) is satisfied, then $S$ in Eq. (\ref{ineq2}) is chosen to make the expression zero. Therefore, the slack variable $S$ must be in the range $0 \le S \le \max_x W - \sum_{i=1}^n w_i x_i$. To represent the slack variable $S$ in binary form, it can be decomposed into $N = \lfloor \log_2(\max_x W - \sum_{i=1}^n w_i x_i)\rfloor+1$ binary variables:

\begin{equation}\label{SB2}
S = \sum_{k=0}^{N-1} 2^k s_k,
\end{equation}
where $s_k$ are the binary variables representing the slack variable. Then, the inequality constraint in Eq. (\ref{ineq2}) can be added to the QUBO formulation as a penalization term given by

\begin{equation}\label{Ineq_EF2}
\lambda_1 \left( W - \sum_{i=1}^n w_i x_i - \sum_{k=0}^{N-1} 2^k s_k\right)^2,
\end{equation}
where $\lambda_1$ is a penalization coefficient that needs to be chosen appropriately to satisfy the inequality constraint. The typical form of the QUBO formulation is given by

 \begin{eqnarray}\label{QUBO_slack}
 \min_x \left(\sum_{i=0, j>i}^{n-1} q_{ij}x_{i}x_{j} + \sum_{i=0}^{n-1} q_{ii} x_i 
 + \lambda_0 \left(\sum_{i=1}^n c_i x_i - C\right)^2\nonumber\right.\\
\left. + \lambda_1 \left( W - \sum_{i=1}^n w_i x_i - \sum_{k=0}^{N-1} 2^k s_k\right)^2 \right).
 \end{eqnarray}
 
\subsection{Ising Hamiltonian}
The last step to represent the QUBO problem as an Ising Hamiltonian is to change the $x_i$ variables to spin variables $z_i \in \{1, -1\}$ by the transformation $x_i = (1 - z_i) / 2$. Note that Eq. (\ref{QUBO_unbalanced}) or Eq. (\ref{QUBO_slack}) can ultimately be rewritten as Eq. (\ref{QUBO_form}) plus a constant value. Hence, Eq. (\ref{QUBO_form}) represented in terms of the Ising Hamiltonian model reads

\begin{equation}\label{IsingH}
H_c(\mathrm{z}) = \sum_{i=0}^{n-1}\sum_{j>i}^{n-1} J_{ij} z_i z_j + \sum_{i=0}^{n-1} h_{i} z_i + \text{offset}
\end{equation} 
where $J_{ij}$ and $h_i$ are real coefficients that represent the combinatorial optimization problem, and the offset is a constant value. Since the offset does not affect the location of the optimal solution, it can be left out for the sake of simplicity.

\subsection{The traveling salesman problem}
The traveling salesman problem (TSP) is one of the most widely studied problems in combinatorial optimization. In this problem, a traveler is given the task to go to a set of $n$ cities passing through each city once and returning to the starting city. The problem is to find the route with the shortest distance. This problem can quickly become very difficult as the number of possible routes grows as $(n-1)!/2$, and the best known classical algorithm that provably solves the TSP has complexity $O(2^nn^2)$~\cite{HeldKarp1962TSP,Bellman1962TSPDynamicProgramming}.

\begin{figure}[htbp]
\includegraphics[width=8.5cm]{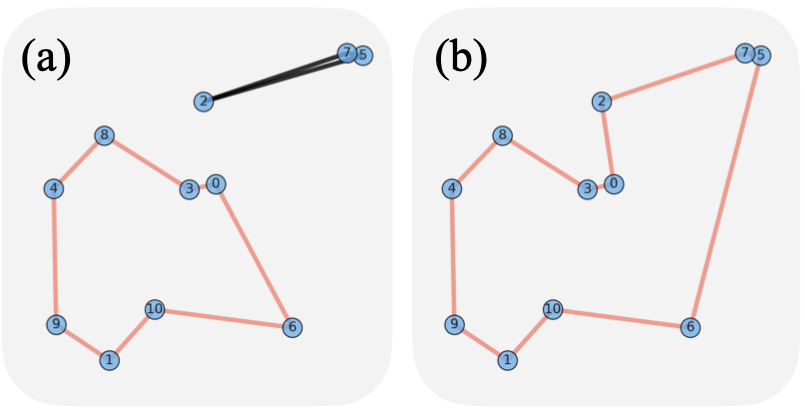}
\caption{TSP problem with 11 cities. (a) the degree LP relaxation of the problem. (b) the optimal solution found after adding the sub-tour $Q = \{2, 5, 7\}$.}
\label{TSP-subtours}
\end{figure}

In this work, we focus on the DFJ formulation of the symmetric TSP, which is known as one of the best classical approaches to solving the TSP in practice. This formulation considers the edges as the variables of the problem. The objective is to find the minimum

\begin{equation}\label{TSP}
\min_x\sum_{i=0, j > i}^{n-1} c_{ij}x_{ij},
\end{equation}
subject to the set of constraints,

\begin{equation}\label{TSPEC1}
\sum_{i=0, i \neq j}^{n-1} x_{\mathrm{sort}(ij)} = 2 \qquad \forall j = 1, ..., n,
\end{equation}
where $n$ is the number of cities (nodes), $c_{ij}$ is the distance between the cities $i$ and $j$, $x_{ij}\in\{0,1\}$ is a binary variable that indicates if the path from the city $i$ to the city $j$ is contained in the final route, $\mathrm{sort}(ij)$ is the sorted indices i, j. Equations (\ref{TSP})--(\ref{TSPEC1}) form the degree LP relaxation of the TSP \cite{cook2006pursuit}. In this model, solutions with sub-tours are still valid and can even be optimal, as shown in Fig. \ref{TSP-subtours} (a), which depicts the optimal solution to an 11-city TSP problem using the relaxation model. However, to avoid sub-tours, additional restrictions are required. One approach is to consider all possible sub-tours in the solution, and these constraints are provided by

\begin{equation}\label{TSPIC}
\sum_{i \in Q} \sum_{j \neq i, j \in Q} x_{ij}\leq |Q| - 1 \ \  \forall Q \subsetneq \{0, ..., n-1\}, |Q| \ge 3,
\end{equation}
where $Q$ represents a sub-tour on a set of cities and $|Q|$ denotes the number of cities in the sub-tour. However, the number of inequality constraints added by this approach grows exponentially with the number of cities, making it infeasible for large instances. Specifically, the number of constraints added by this approach is $2^{n/2} - 2n - 2$ \cite{Khumalo2022}.

To overcome this issue, an alternative method called "finger in the dike" has been proposed \cite{Dantzig1954}. This method adds inequality constraints for the sub-tours as they appear in the solution, resulting in a smaller set of constraints. Specifically, in each iteration, we identify the shortest sub-tour in the current solution and add an inequality constraint for that sub-tour. This process continues until a valid tour is obtained. For instance, as illustrated in Fig. \ref{TSP-subtours}, we first obtain the solution of the degree LP relaxation, which results in the solution shown in Fig. \ref{TSP-subtours} (a). Then, we add a constraint on the sub-tour $Q = \{2, 5, 7\}$ in a new iteration of the problem, which leads to the solution shown in Fig. \ref{TSP-subtours} (b). Finally, the algorithm terminates when a valid tour is found.

We propose a variation of the "finger in the dike" method, which considers the fact that multiple suboptimal solutions can be found with or without sub-tours on an iteration. Our approach aims to strike a balance between reducing the number of constraints and directing the search towards finding solutions without sub-tours. The algorithmic details of our proposed approach are described below.

\begin{algorithm}[!ht]
\DontPrintSemicolon
  
  \KwInput{Distance between cities $c_{ij}$ for $n(n-1)/2$ edges $x_{ij}$ of $n$ cities. QUBO parameters $\lambda$. The maximum number of iterations $N$.}
  \KwOutput{A bitstring of the variables $x_{ij}$ with 1 indicating a selected edge and 0 otherwise for the best solution found.}
  \KwData{Position (x, y) for each city.}
  
  min\_distance $\gets$ Big number
  
  subtour\_constraints $\gets \{\}$ 
  
  \For{k=0; k++; $k < N$}
    {
     	get QUBO formulation of degree LP relaxation
     	
     	add subtour\_constraints to QUBO formulation

	get a set of bitstrings $\{\mathrm{\bar x}\}$ using the optimization method, e.g, quantum annealing or CPLEX
	
	\For{$\bar x\in\{\mathrm{\bar x}\}$}{
	 \If{Eq. (\ref{TSPEC1}) is True for $\mathrm{\bar x}$}{
	 	\If {$\mathrm{\bar x}$ includes no sub-tours}
			{add $\mathrm{\bar x}$  to the solutions
			
			current\_distance $\gets$ distance($\mathrm{\bar x}$)
			
			min\_distance $\gets$ min(current\_distance, min\_distance)
			}
		\Else
			{
			add $\mathrm{\bar x}$  to relaxation\_solutions
			}
	}
	}

    subtour\_constraints $\gets \{\}$ 
    
	\For{$\mathrm{\hat x}$ in relaxation\_solutions}
	{
	\If{distance($\mathrm{\hat x}$) $\leq$ min\_distance}
	{
	add the sub-tour from $\mathrm{\hat x}$ with the smallest number of cities to subtour\_constraints
	}
	}

    }
    
\caption{Sub-tour solutions elimination. The finger in the dike method.}
\end{algorithm}
 
\subsection{Solvers}

We use different solvers in our study to compare the slack variables and unbalanced encoding using classical and quantum methods. For classical methods, we use simulated annealing and branch and bound. For quantum methods, we use quantum annealing and the D-Wave hybrid method to obtain solutions for various TSP problems. We give a brief description for each of the methods used.

Simulated annealing is a metaheuristic algorithm used for finding the global optimum of a given problem \cite{kirkpatrick1983optimization}. Like many algorithms, it has several parameters that need to be set appropriately for the algorithm to work efficiently. We use the simulated annealing algorithm implemented in the D-Wave Ocean SDK \cite{dwave-docs}. We leave the default parameters of the inverse temperature $\beta \in \{0.09, 9.6\}$ with a geometric cooling schedule and 1000 sweeps.

Branch and bound is a technique used in optimization problems to divide the problem into smaller sub-problems, and the solutions to these sub-problems are used to narrow the search space, leading to a faster convergence towards the optimal solution. We use the branch and bound method from CPLEX \cite{cplex}, which is a commercial optimization software developed by IBM that is widely used to solve a variety of optimization problems. This software uses the branch and bound algorithm along with cutting planes techniques to solve LP problems such as the TSP. We refer to this method as CPLEX reference in what follows.

The quantum annealer used is the D-Wave Advantage 5.3 System JUPSI, a quantum computing system developed by D-Wave Systems Inc that has more than 5000 qubits located in J\"ulich, Germany. It is designed to solve optimization problems using quantum annealing, a process that exploits quantum mechanical effects to find low energy states of Ising Hamiltonians.

Finally, the hybrid method is a quantum-classical algorithmic approach that leverages both classical and quantum computing resources to solve optimization problems. By combining classical optimization algorithms with quantum annealing, the hybrid method enhances the efficiency and precision of the solution, surpassing the capabilities of quantum annealing alone.

\section{Results}\label{Sec:Results}
The results presented in this section are for random problems ranging from 6 to 45 cities, the coordinates (x, y) for each of the cities are selected randomly from $[-1, 1]$. The cost of travelling between two cities is always given by the two-dimensional Euclidean distance. 

The parameters $\lambda_{0,1,2}$ in Eq.~(\ref{QUBO_unbalanced}) were tuned using a random case with 12 cities. The procedure consists of three parts. First, we obtain the sub-tour constraints (i.e., 4 in this particular case) by running Algorithm 1 using CPLEX. Then an Ising Hamiltonian model,
based on the current $\lambda_{0,1,2}$ parameters, is generated and solved using the D-Wave Advantage solver with 5000 samples. Finally, from the resulting bitstring solutions, only the ones that satisfy the given constraints are selected. The mean energy of these feasible solutions is then calculated as the objective function value. To find the minimum of this objective function, the Constrained Optimization BY Linear Approximation (COBYLA) solver \cite{Powell1994COBYLA} is employed. The results of this optimization method starting with an initial guess $\lambda_{0,1,2} = \{1, 1, 0.1\}$ were $\lambda_{0,1,2} = \{0.88, 0.46, 0.54\}$. 
Since $0.88$ was found to be a good parameter to respect the equality constraints, we use the same value also for constraints in the slack variables approach, i.e., $\lambda_{0,1} = \{0.88, 0.88\}$ in Eq.~(\ref{QUBO_slack}).

\subsection{Quantum Annealer: D-Wave Advantage}

\begin{figure}[!tbh]
\centerline{\includegraphics[width=8.5cm]{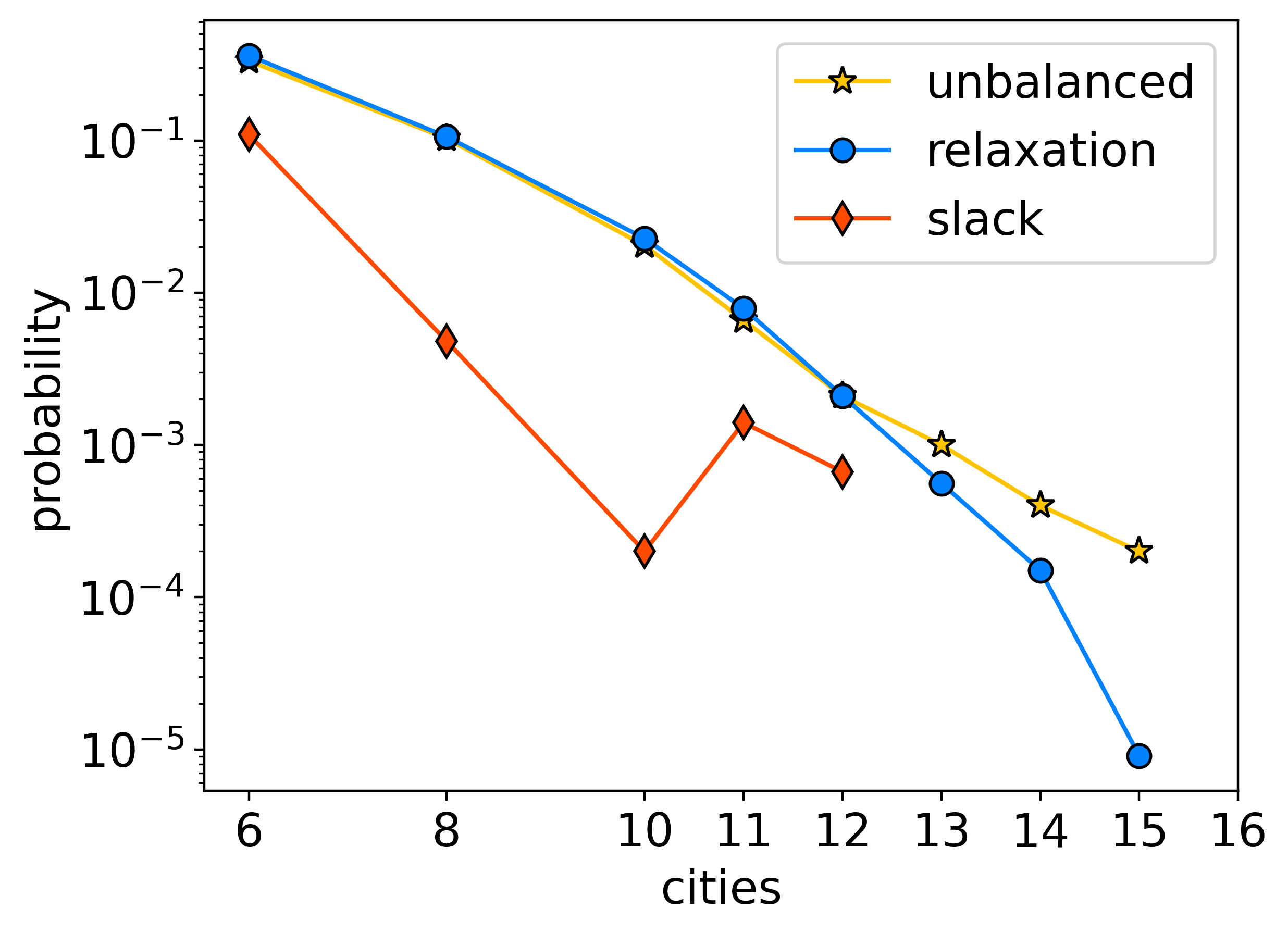}}
\caption{Probability of finding a valid solution on the QPU using the different methods, the unbalanced penalization method (yellow stars), the degree LP relaxation (blue circles), and the slack variables approach (red diamonds). Lines are guides to the eye.}
\label{probability-qpu}
\end{figure}

Figure \ref{probability-qpu} shows the probability of finding valid solutions for various TSP problems using the D-Wave Advantage. Our experiments involved testing three different scenarios: the degree LP relaxation, unbalanced penalization, and slack variables encodings. We excluded the results for the 7 and 9-city problems because the degree LP relaxation consistently identified the optimal solution without any additional inequality constraints.

Using the slack variables approach for inequality constraints, we were able to identify solutions for up to 12 cities. However, for problems beyond that size, we were unable to find any valid solutions. This is because adding a sub-tour constraint to the set of inequality constraints requires more qubits and connections on the actual device, as shown in Table \ref{tab1}. For instance, the slack encoding for 12 cities needs 68 logical qubits and 676 connections, which is similar to the degree LP relaxation. However, for 13 cities, the requirements for connections are 74\% larger than the degree LP relaxation, which can be challenging on real hardware. In addition, due to the problem of embedding \cite{Cai2014DWaveEmbedding}, the number of physical qubits used on the device are even larger, since the connectivity between the physical qubits (which is 15 on the Pegasus topology) is usually not enough to represent all connections between the binary problem variables.

Our experiments revealed that the degree LP relaxation and the unbalanced penalization method exhibit similar performance for TSP problems with up to 12 cities. However, beyond this point, the probability of finding valid solutions using the degree LP relaxation drops considerably (see Fig. \ref{probability-qpu}). This can be attributed to the fact that for small TSP problems, there are relatively few sub-tours in the low-energy bitstrings, making the degree LP relaxation a suitable option only for small TSP problems. Table \ref{tab1} shows that the two methods have similar requirements, except when sub-tours with more than 3 cities are found. In such cases, the penalization term for the sub-tour inequalities increases the number of connections. In the event that a three-city sub-tour is identified, the application of the aforementioned inequality constraint does not necessitate additional connections. This is due to the fact that the equality constraints already introduced the required connections.

\begin{table}[htbp]
\caption{Qubit and connection requirements of the different problem encodings.}
\begin{center}
\begin{tabular}{|c|c|c|c|c|c|c|c|c|}
\hline
\textbf{}&\multicolumn{8}{|c|}{\textbf{Cities}} \\
\cline{2-9} 
\textbf{} & \textbf{\textit{6}}& \textbf{\textit{8}}& \textbf{\textit{10}}& \textbf{\textit{11}} & \textbf{\textit{12}} & \textbf{\textit{13}} & \textbf{\textit{14}}  & \textbf{\textit{15}}\\
\hline
\multicolumn{9}{|c|}{\textbf{LP relaxation}} \\
\hline
qubits &15 & 28 &  45 & 55 & 66 &  78 & 91 & 105\\
\hline
connections & 60 & 168 & 360 & 495 & 660 &  858 & 1092 & 1365 \\
\hline
\multicolumn{9}{|c|}{\textbf{Unbalanced}} \\
\hline
qubits &15 & 28 &  45 & 55 & 66 &  78 & 91 & 105\\
\hline
connections & 60 & 168 & 360 & 498 & 678 &  858 & 1092 & 1575 \\
\hline
\multicolumn{9}{|c|}{\textbf{Slack}} \\
\hline
qubits &17 & 33 &  50 & 57 & 68 &  81 & 93 & 108\\
\hline
connections & 67 & 223 & 415 & 502 & 676 &  1155 & 1108 & 1662\\
\hline
\end{tabular}
\label{tab1}
\end{center}
\end{table}

\begin{figure}[!tbh]
\includegraphics[width=8.5cm]{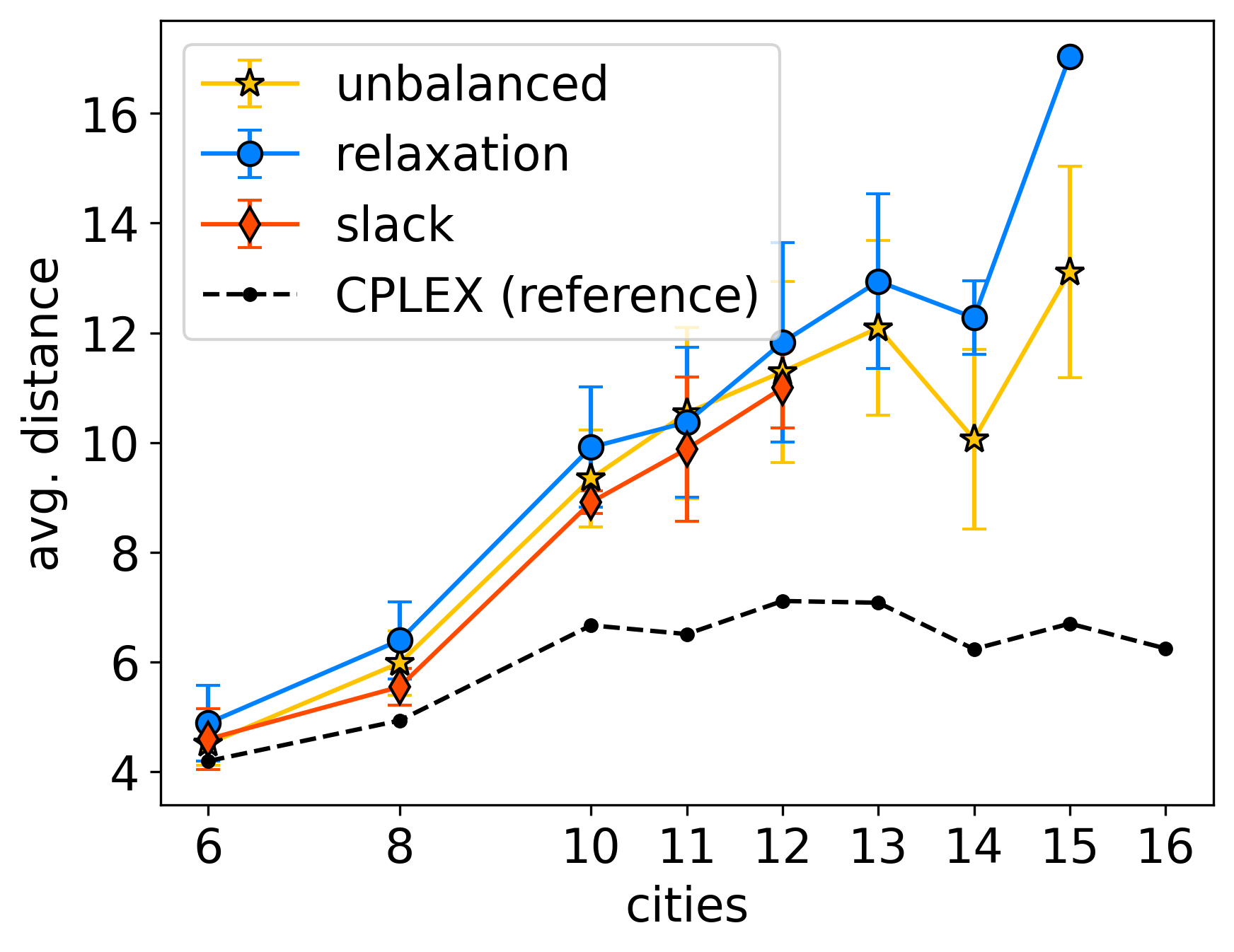}
\caption{Average tour distance of the valid solutions found using the different methods. The error bars indicate one standard deviation of the distances. The markers are the same as in Fig.~\ref{probability-qpu}. For reference, the black points are the results of the CPLEX solver. Lines are guides to the eye.}
\label{mean-qpu}
\end{figure}

Figure \ref{mean-qpu} displays the average distance of solutions obtained through the degree LP relaxation, unbalanced penalization and slack variables encodings, and including the CPLEX solver's solution. For the CPLEX solver, we utilized the LP problem, which avoids translating the problem into the QUBO formulation with penalization terms for the constraints. Instead, it employs a branch-and-bound approach, which is effective for solving TSP instances of moderate sizes.

\begin{figure}[!tbh]
\includegraphics[width=8.5cm]{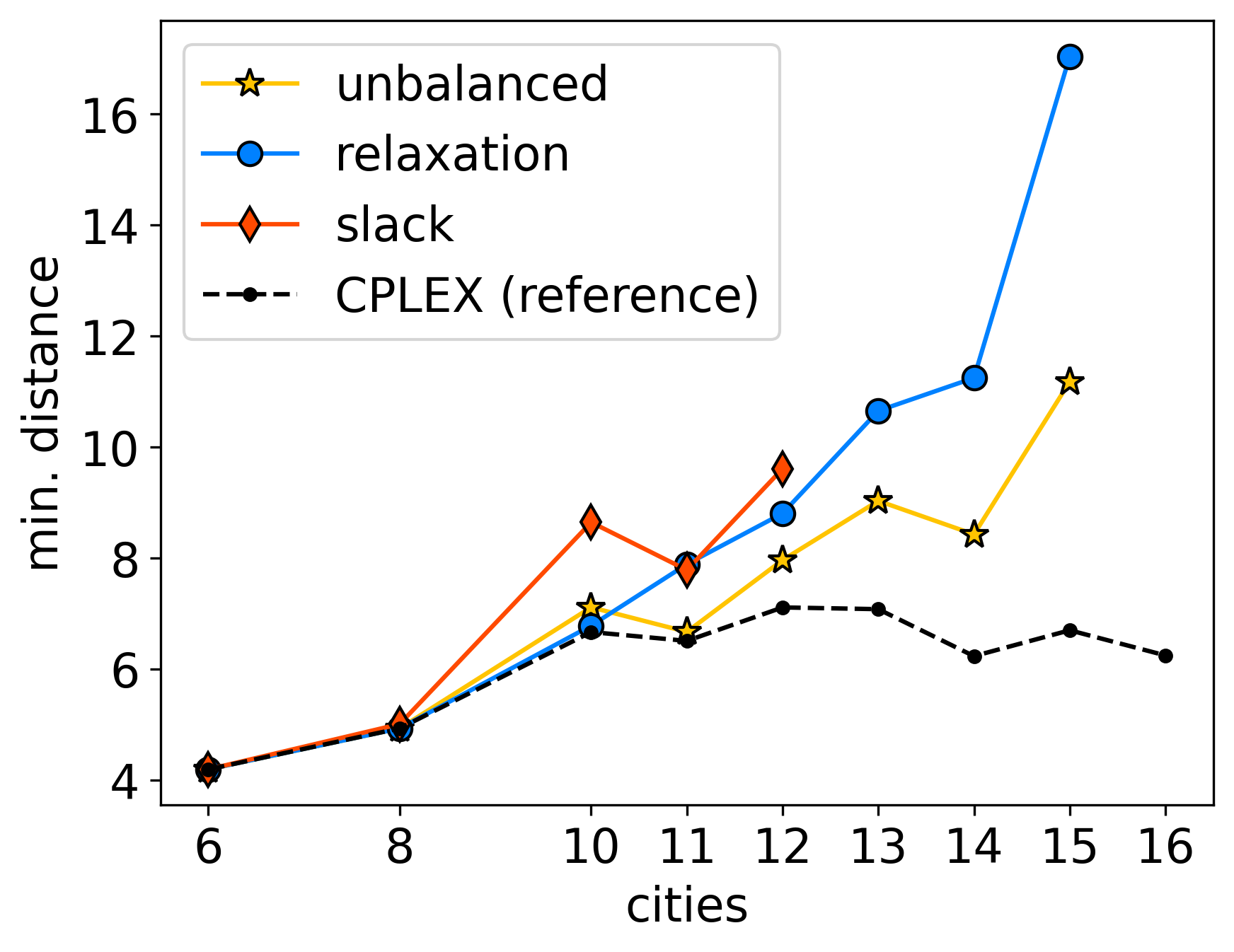}
\caption{Minimum distance of the valid solutions found using the different methods. The markers are the same as in Fig.~\ref{mean-qpu}.}
\label{min-qpu}
\end{figure}

Figure \ref{min-qpu} depicts the minimum distance achieved through various methods. For TSP problems involving up to 8 cities, all three methods found the optimal solution. However, as the number of cities increased beyond 10, the unbalanced penalization method proved to be more effective in finding shorter paths. In these cases, the error bars represent the standard deviation of all valid solutions discovered.

\subsection{Hybrid Solver}
\begin{figure}[!tbh]
\includegraphics[width=8.5cm]{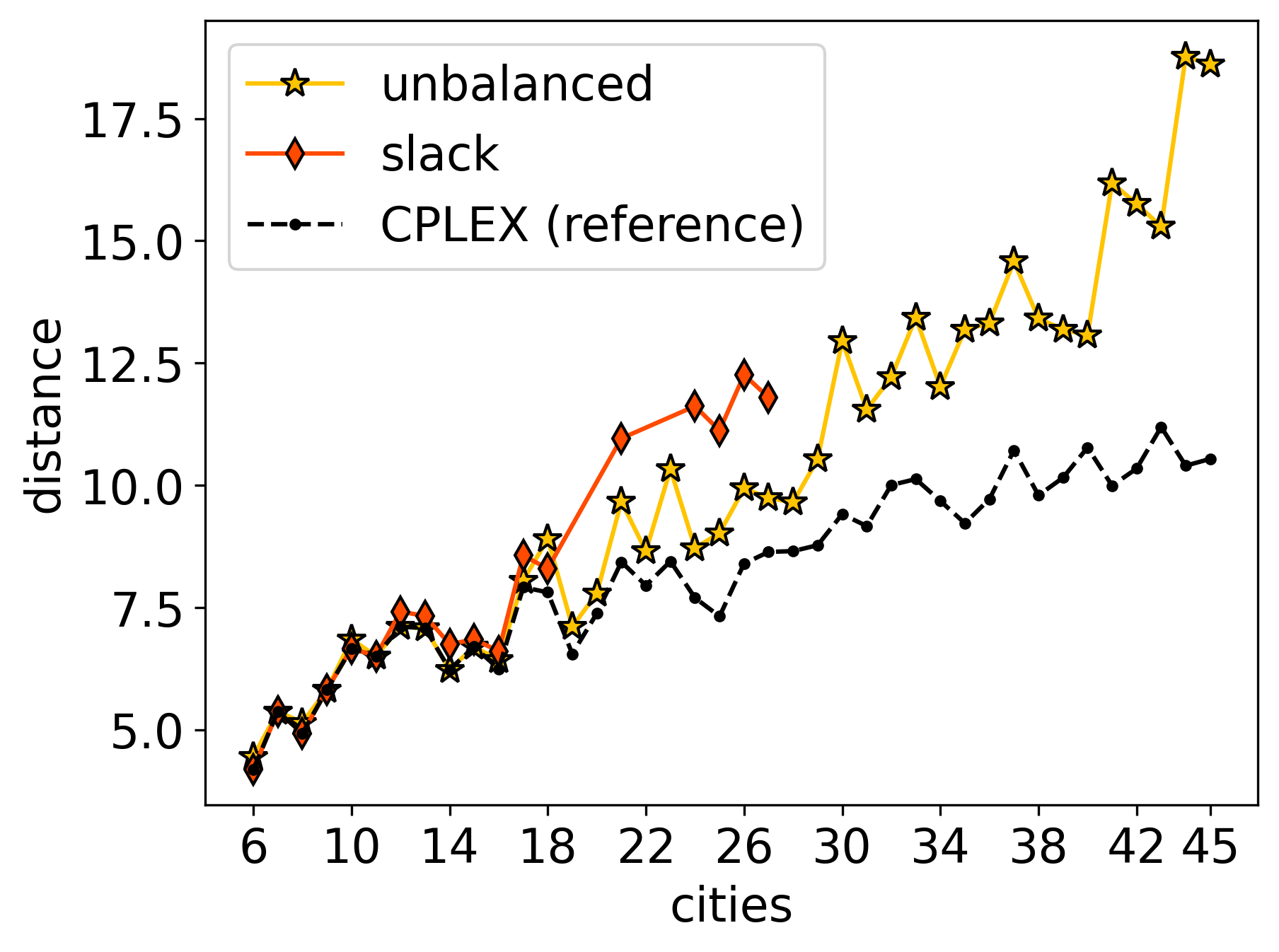}
\caption{Comparison of the unbalanced penalization method (yellow stars) and the slack method (red diamonds) using the D-Wave hybrid solver. For reference, black points represent the result of the CPLEX solver. Lines are guides to the eye.}
\label{hybrid}
\end{figure}

The D-Wave hybrid solver employs a combination of quantum and classical resources to tackle complex problems. In Fig. \ref{hybrid}, we present the outcomes obtained through the D-Wave hybrid solver for various TSP random problems. By utilizing the slack variables approach, solutions are comparable to the unbalanced penalization method for up to 18 cities. However, the hybrid solver struggles to find solutions beyond this point, and the tour distance is often longer than that of the unbalanced method. Additionally, using the slack variables encoding, the hybrid solver fails to discover any solutions beyond 27 cities. In contrast, the unbalanced penalization approach consistently yields valid solutions for all tested cases, up to 45 cities. This unexpected result suggests that when using the unbalanced penalization method, many of the low-energy bitstrings still represent valid solutions.

\subsection{Unbalanced penalization using different solvers}

\begin{figure}[!tbh]
\centerline{\includegraphics[width=8.5cm]{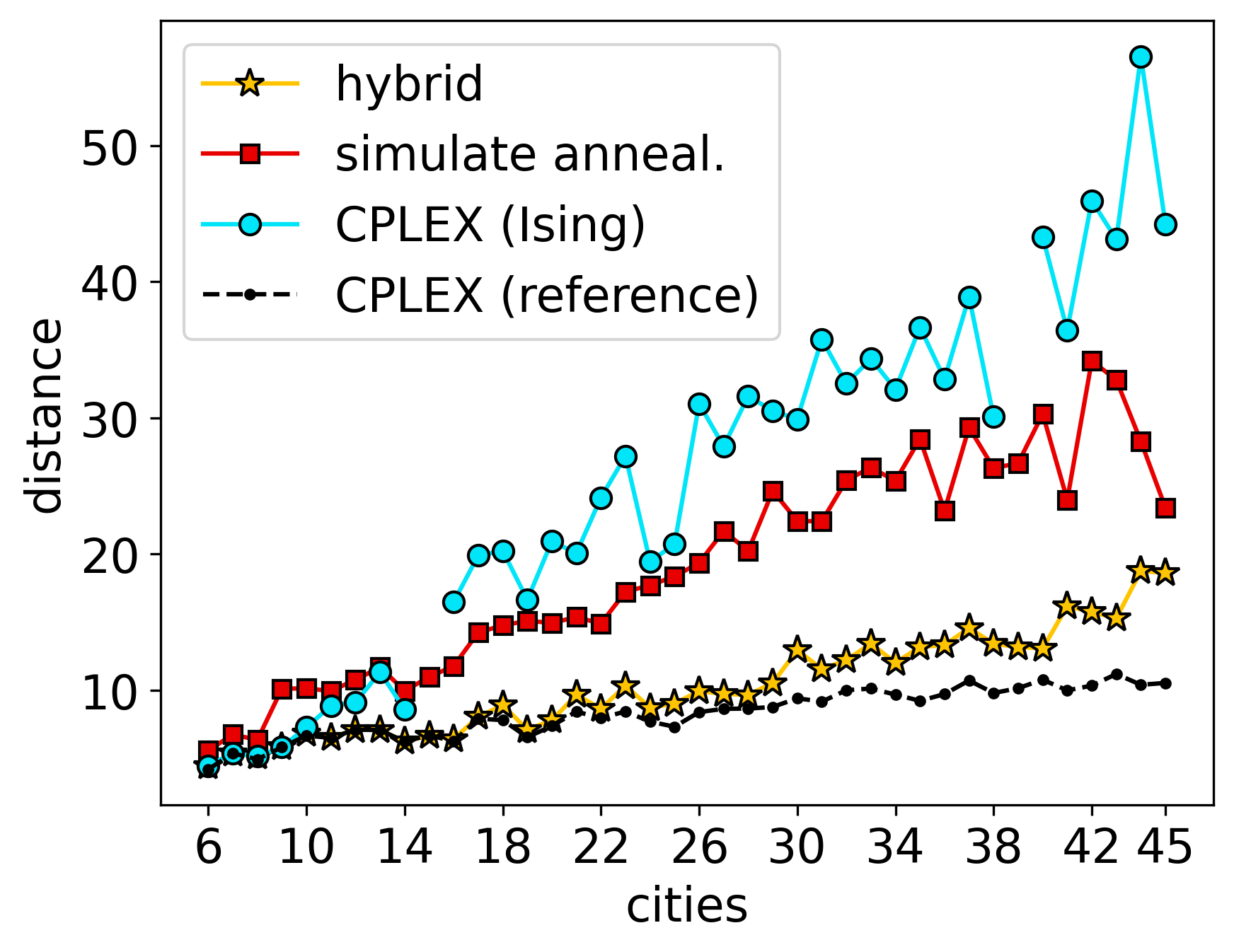}}
\caption{Performance of the unbalanced penalization method using different solvers, namely the D-Wave hybrid solver (yellow stars), simulated annealing (red squares), and the CPLEX solver (turquoise circles) using the Ising formulation . For reference, black points represent the result of the CPLEX solver using the LP formulation. Lines are guides to the eye.}
\label{diffsolv}
\end{figure}

The results of the unbalanced penalization method applied to the TSP problem using different solvers are presented in Fig. \ref{diffsolv}.
The maximum time for finding a solution is set at 6 seconds for CPLEX and 3 seconds for the hybrid solver. Each problem is executed 10 times, and the solution that fulfills all constraints with the minimum distance found is recorded. Notably, when CPLEX is used to solve the QUBO or Ising formulation of the problem (turquoise circles), the performance of CPLEX decreases significantly and sometimes a solution is not found after the 10 executions.

In our experiments, the hybrid solver outperformed all other solvers using the Ising Hamiltonian representation of the problem. Notably, all solutions presented here utilized the unbalanced penalization method, and in almost all cases, valid solutions were obtained. This suggests that the unbalanced penalization method performs well and can identify valid solutions for the TSP problem even using classical solvers.

\section{Conclusions}\label{Sec:Conclusions}

In this work, we have demonstrated the effectiveness of the unbalanced penalization method for encoding inequality constraints of combinatorial optimization problems using the TSP as a case study. We have tested the method on TSP problems ranging from 6 to 45 cities and compared its performance with that of the slack variables approach. Our experiments were conducted on various solvers including the D-Wave Advantage QPU, the D-Wave hybrid solver, CPLEX using both the LP representation and the QUBO or Ising Hamiltonian representation, and simulated annealing.

Our findings show that the unbalanced penalization approach outperforms the slack variables approach, and we were able to find solutions for larger TSP problems than what has been previously reported in the literature for QPUs (8 and 12 cities) \cite{Jain2021, Gawalleck2022}. We believe that solving combinatorial optimization problems is essential in benchmarking new quantum algorithms and technologies. Therefore, we expect more effort to be devoted to using combinatorial optimization problems, such as the TSP, to assess the progress with current and future devices.

An advantage of the unbalanced penalization approach is that it does not add further connectivity requirements if the sub-tours have only three cities. Since the number of connections is a limiting factor in quantum devices, reducing this requirement is crucial. Therefore, looking for algorithms that can reduce the number of required connections is an essential step in advancing this technology.

We also show that unbalanced penalization is a good choice to encode inequality constraints when using classical solvers. Even though we cannot  be sure that the optimal solution is encoded in the ground state, the method still yields many valid sub-optimal solutions, which may be advantageous in practice.

One last important observation is that the performance of the CPLEX solver decreases considerably when the problem is presented as a QUBO, while for the LP formulation, it takes an insignificant amount of time to find the optimal solution even for large numbers of cities. For the QUBO formulation, CPLEX takes much longer and from all the methods tested this was the one that gave the worst results. The quantum annealer, in contrast, gives much better solutions in a short time, for both the hybrid solver and the QPU, though the latter only found valid solutions up to 15 cities.

\section*{Acknowledgment}

J. A. Monta\~nez-Barrera and P. H. van den Heuvel acknowledge support by the German Federal Ministry of Education and Research (BMBF), funding program Quantum technologies - from basic research to market, project QSolid (Grant No. 13N16149). D.~Willsch acknowledges support from the project J\"ulich UNified Infrastructure for Quantum computing (JUNIQ) that has received funding from the German Federal Ministry of Education and Research (BMBF) and the Ministry of Culture and Science of the State of North Rhine-Westphalia. The authors gratefully acknowledge the Gauss Centre for Supercomputing e.V. (www.gauss-centre.eu) for funding this project by providing computing time on the GCS Supercomputer JUWELS at Jülich Supercomputing Centre (JSC).

\bibliographystyle{plain}
\bibliography{main}

\end{document}